\documentclass[onecolumn,showpacs,floatfix,10pt,nofootinbib]{revtex4}
\pdfoutput=1
\usepackage{epsfig}\usepackage{float}
\RequirePackage{amssymb,amsmath}
\usepackage{makeidx}
\usepackage[latin1]{inputenc}
\usepackage{graphicx}
\usepackage{indentfirst}
\usepackage{color}
\begin{document}

\title{Production of isomeric states in the deuteron-induced reaction of gold at incident energy 4 GeV}
\author{A. R. Balabekyan$^a$, N. A. Demekhina$^b$, G. S. Karapetyan$^c$, D. R. Drnoyan$^d$, V. I. Zhemenik$^d$, J. Adam$^d$, L. Zavorka$^d$, A. A. Solnyshkin$^d$, V. M. Tsoupko-Sitnikov$^d$}
\affiliation{a) Yerevan State University \\ A. Manoogian, 1, 025, Yerevan, Armenia \\ b) Yerevan Physics Institute, Alikhanyan Brothers 2, Yerevan 0036, Armenia\\ Joint Institute for Nuclear Research (JINR), Flerov Laboratory of Nuclear Reactions (LNR), Joliot-Curie 6, Dubna 141980, Moscow region Russia \\ c) Instituto de Fisica, Universidade de S\~ao Paulo \\ Rua do Matao, Travessa R 187, 05508-900 S\~ao Paulo, SP, Brazil \\ d) Joint Institute for Nuclear Research (JINR), Flerov Laboratory of Nuclear Reactions (LNR), Joliot-Curie 6, Dubna 141980, Moscow region Russia}

\begin{abstract}
The independent cross section ratio for production of nuclei from $^{197}$Au targets irradiated with 4 GeV deuterons have been measured by off-line $\gamma$-spectroscopy. On the basis of the measured independent cross section ratio of $^{198m, g}$Au the average intrinsic angular momentum of the primary nucleus was estimated by means of a simple statistical-model analysis based on the formalism developed by Huizenga and Vandenbosch.
\end{abstract}
\pacs{}
\maketitle

\section*{1 Introduction}
The search for alternative, innovative and efficient sources of energy was an advantage branch of the research of physicists during the decade of last centure. One of the new methods to control the release of energy stored in an atom's nucleus was a production of a nuclear isomers which considered as most of promising and attractive way for the production of powerful sources of soft $\gamma$-ray radiation in the form of excitation energy of isomeric state \cite{Karamian}. An isomer is a long-lived excited state of an atom's nucleus, a state in which decay back to the nuclear ground state is inhibited. The nucleus of an isomer thus holds an enormous amount of energy which is released when the nuclei transition from a high energy state to a lower one.

Nuclear isomers play an important role in medicine applications due to soft $\beta$- and $\gamma$-radiation and the absence of radioactive contamination after its decay. Isomers, which have suitable half-lifes and radioactive properties, can be successfully used as radioactive sources for therapy of different types of cancer.

The methods to product such isomers are different, for example, the chemical separation.
However, sometimes due to a large amount of impurities, it is not possible to distinguish given isomer using this method. Identification of the different spin states is often performed by induced activity method. Such method allows to register the reaction products with nuclear characteristics convenient for measurements and determine the cross section of these products after accumulation in a nuclear reaction. 

The population of the ground ($\sigma_{g}$) and metastable  ($\sigma_{m}$) states of a nucleus depends on the angular momentum in the entrance channel of the reaction, excitation energy of the residual nucleus, and the type of particles emitted during its de-excitation. Study of the angular momenta of the rection fragments can provide insight into an information about the configuration of the nuclear system at high excitation. The information about the primary angular momenta of the fragments can be obtained from the measurements of independent isomer ratios of the reaction products. Usually, such measurements are based on the cross section ratio of high-spin ($I_{h}$) state to low-spin ($I_{l}$) state (isomer ratios, $IR = \sigma(I_{h})/\sigma(I_{l}$). 

Above mentioned technique imposes also some restrictions on the measured independent cross sections due to the contribution from the $\beta^{\pm}$ decays of neighboring unstable isobars, i.e. the cumulativity of their production. Thus, only a few part of the experimental data can be used to determine the characteristics of the reaction under study.

De-excitation of primary products take places by the emission of prompt neutrons and $\gamma$-rays until populating the final state level with different spins. Neutrons and $\gamma$-rays carry away different amount of energy and angular momentum, hence, change the initial distribution of the primary products. As a result, the primary products originating from reaction remnants, have a wide range of angular momenta and excitation energies. The comparison of the measured isomer cross section ratios with calculations within different statistical models \cite{Zhuikov, Patronis} helps to estimate the spin associated with the reaction products, produced in different exit channels. 

Measurements of of high-state isomers are an experimental challenge because they are difficult to measure, therefore the experimental information is very scarce up to now. There are some experimental data with photons \cite{Demekhina}, protons \cite{Zhuikov, Orth} and heavy-ions \cite{Aumann} at intermediate energy range. However, there is no measured independent isomer-yield ratios for gold target at high energies have been reported.

In this work the independent cross section ratios for the reaction products on gold target were determined at the energy of deuterons 4.4 GeV measured using induced activity method. The interpretation of the high-spin states population may be important for the understanding of the mechanism of intermediate- and high-energy particle interaction with nuclei.

\section*{2 Experimental Procedure}

Beam of 4 GeV deuteron from the Nuclotron of the VBLHEP, JINR was used to irradiate gold target of thickness
39.13 mg/cm$^{2}$, surrounded by Mylar catcher foils of the same size.  The Mylar foils in contact with the gold served as forward and backward catchers. The irradiated target foils without chemical separation were measured using a HpGe spectrometers (with detectors energy resolution on average 0.23\% at an energy of 1332 keV). The energy-dependent detection efficiency of the HpGe detectors was determined with standard calibration sources of $^{22}$Na, $^{54}$Mn, $^{57;60}$Co, $^{137}$Cs,  $^{154}$Eu, $^{152}$Eu, and $^{133}$Ba. Nuclear properties, used for identification of observed isotopes as nuclear transition energies, intensities, and half-lives were taken from literature \cite{Firestone}. The experiment in question was described in detail elsewhere \cite{Balabekyan}.

The isomer cross section rations can be calculated from measured activities of the metastabil and ground states of the product nuclei. Ground-state nucleus can be formed directly from the target nucleus and/or indirectly through the decay of the metastabil-state nucleus. The production of the given isomeric pair and its decay during the activation time, $t_{i}$, can be described by the following differential equation:

\begin{eqnarray}
\frac{dN_{m}}{dt} = \sigma_{m} - \lambda_{m}N_{m}
\end{eqnarray}

\begin{eqnarray}
\frac{dN_{g}}{dt} = \sigma_{g} - \lambda_{g}N_{g} + p\lambda_{m}N_{m}
\end{eqnarray}
\noindent 
where $N_{i}$ is the numbers of nuclei for i = (m, g) state, $\lambda_{m}$ and $\lambda_{g}$ are the decay constants of these states, and $p$ is the branching ratio for the decay of metastable to ground state.

Solving equations (1) and (2) in the three time intervals (the irradiation time $t_{1}$; the time of exposure between the end of the irradiation and the beginning of the measurement $t_{2}$; the time of measurement$t_{3}$), the isomer ratio can be derived from formula, which involves the ratio of the areas under the peaks of measured $\gamma$-transitions \cite{Vanska, Kolev}:

\begin{widetext}
\begin{eqnarray}
\frac{\sigma_{m}}{\sigma_{g}} = \left[\frac{\lambda_{g}(1-e^{\lambda_{m}t_{1}})e^{\lambda_{m}t_{2}}(1-e^{\lambda_{m}t_{3}})}{\lambda_{g}(1-e^{\lambda_{m}t_{1}})e^{\lambda_{m}t_{2}}(1-e^{\lambda_{m}t_{3}})} 
\times\left(\frac{k_{m}N_{g}\eta_{m}\epsilon_{m}}{k_{g}N_{m}\eta_{g}\epsilon_{g}} - p\frac{\lambda_{g}}{\lambda_{g}-\lambda_{m}}\right) + p\frac{\lambda_{m}}{\lambda_{g}-\lambda_{m}}\right]^{-1},
\end{eqnarray}
\end{widetext}

Here, $N$ is the area under the photopeak with energy $E_{\gamma}$; $\lambda$ is the decay constant (min$^{-1}$); $\eta$ is the intensity of $\gamma$-transitions; $k$ is the total coefficient of $\gamma$-ray absorption in target and detector materials, and $\epsilon$ is the $\gamma$-ray detection efficiency and $p$ is the contribution from the metastable state to the ground state. The subscripts $m$ and $g$ label variables referring to the metastable and the ground states, respectively. Below the schemes of the determination the ground and metastable sates of the products under study are shown:

\begin{itemize}
\item The cross section of isomeric state $^{95}$Nb$^{m}$ ($T_{1/2}$=86.64 h) at the energy $E_{\gamma}$=235.68 keV, was calculated by a formula from \cite{Deppman} that takes into account the contribution $f_{ij}$=0.0088 of parent isotope $^{95}$Zr ($T_{1/2}$=64.02 d). The cross section of ground state $^{95}$Nb$^{g}$ ($T_{1/2}$=34.98 d) at the energy $E_{\gamma}$=766.0 keV, was calculated as an independent cross section with the contribution $f_{ij}$=0.991 of parent isotope $^{95}$Zr ($T_{1/2}$=64.02 d). The isomer ratio was calculated by formula as an independent cross section (3) with contribution from the metastable state to the ground state $p$=0.976. 
\item The cross section of isomeric state $^{95}$Tc$^{m}$ ($T_{1/2}$=61.0 d) at the energy $E_{\gamma}$=204.11 keV, was calculated by a formula from \cite{Deppman} as a cumulative cross section. The cross section of ground state $^{95}$Tc$^{g}$ ($T_{1/2}$=20.0 h) at the energy $E_{\gamma}$=765.79 keV, was calculated as a cumulative cross section. Since the contribution of parent isotope $^{95}$Ru is the same for both nucleus of isomeric pair, it was valid to calculate isomer ratio in such case. The isomer ratio was calculated by formula (3) with contribution from the metastable state to the ground state $p$=.
\item The cross section of isomeric state $^{102}$Rh$^{m}$ ($T_{1/2}$=2.9 y) at the energy $E_{\gamma}$=475.1 keV, was calculated by a formula from \cite{Deppman} as an independent cross section with taking into account the contribution $f_{ij}$=0.42 of isotope $^{144}$Pm ($T_{1/2}$=363.0 d). The cross section of ground state $^{102}$Rh$^{g}$ ($T_{1/2}$=207.0 d) at the energy $E_{\gamma}$=556.41 keV, was calculated as an independent cross section. The isomer ratio was calculated by formula (3) with contribution from the metastable state to the ground state $p$=.
\item The cross section of isomeric state $^{184}$Re$^{m}$ ($T_{1/2}$=169.0 d) at the energy $E_{\gamma}$=216.55 keV, was calculated as an independent cross section by a formula from \cite{Deppman}. The cross section of ground state
$^{184}$Re$^{g}$ ($T_{1/2}$=38.0 d) at the energy $E_{\gamma}$=792.07 keV and 903.28 keV, was calculated. The isomer ratio was calculated by formula (3) as an independent cross section with contribution from the metastable state to the ground state $p$=.
\item The cross section of isomeric state $^{198}$Au$^{m}$ ($T_{1/2}$=2.27 d) at the energy $E_{\gamma}$=214.84 keV, was calculated by a formula from \cite{Deppman} as an independent cross section with taking into account the contribution $f_{ij}$=0.86 of isotope $^{97}$Ru ($T_{1/2}$=2.9 d). The cross section of ground state $^{198}$Au$^{g}$ ($T_{1/2}$=2.69 d) at the energy $E_{\gamma}$=411.8 keV, was calculated as an independent cross section with the contribution $f_{ij}$=1.0 of parent isotope $^{198}$Au$^{m}$. The isomer ratio was calculated by formula (3) with contribution from the metastable state to the ground state $p$=.
\end{itemize}

\section*{3 Results and Discussion}

Table I lists cross sections of several isomeric pairs were been successfully detected. The production cross sections of all isotopes were determined as independent yield after subtraction of the contribution from EC and $\beta$-decays.
The main sources of the uncertainties for the present results were estimated due to statistical error, the detection efficiency error, the geometry uncertaintes of the sample position for irradiation and measurement, as well as the errors of the nuclear data used in the calculation such as half-live of the radioactive isotopes and $\gamma$-ray intensities.

As it is well known, the fragments in high energy nuclear collisions are produced by spallation, fission, and multifragmentation. At peripheral collisions which usually take place at relativistic velocities of the projectiles, such processes can be described using the two-step ``abrasion-ablation" model \cite{Hufner} in which
nucleons are removed from the projectile and target nuclei (abrasion), leaving behind highly excited projectile-like and
target-like prefragments with excitation energy $E^{*}$. The abrasion phase is viewed in terms of the ``participant-spectator" system in which the majority of the projectile-target interaction occurs within the participant region of overlap between the colliding nuclei. Prefragments may then evaporate nucleons and other light particles (ablation), losing energy. Reaction residuals are formed in different spin states. The angular momentum of the final nucleus is a result of the transfer of orbital angular momentum from the incident particle/nucleus plus its coupling with the spin of the target nucleus. According to the J. Hufner \cite{Hufner} spallation is the process in which only one heavy fragment with mass close to the target mass $A_{T}$ is formed (the special case of spallation is so-called deep spallation where $M$ is also $M=1$ but $50 < A < 2A_{T}/3$); fission is the process which leads to two heavy fragments in the interval around $A = A_{T}/2$; and multifragmentation is the process whose result is the formation of several (more than two) fragments with $A < 40-50$ amu.

The variation of $IR$ with product mass $A$ is shown in Fig.1 using a logarithmic scale. A dashed curve represents the trend of the data. From the Fig.1 it can be seen that $IR$ shows a rapid increase with increasing mass of the product until about $A \sim 60-65$ amu, where approximately fission process starts to contribute to the residual mass-yeild. With further mass rise, going into the deep spallation region and after up to the near-target product region, the $IR$ values level off. Such behavior suggests that the probability of the population of high-spin state depends on the fragment mass in the fragmentation region of mass-yield distribution. It gradually reaches a plateau for fission-like/deep spallation and as well as spallation product masses. The saturation of isomer ratios in the heavy fragment masses in this work can be explained by increasing the cross section of light fragments at high energy range and competition of different reaction channels such as fragmentation and multifragmentation. In this energy range the energy transfer to the after cascade remnants promotes the new reaction channel opening.

Additinally, as it can be noted from Table I, that in the plateau range the difference of high- and low-spin states ($I_{h} - I_{l}$) increases from 4 $\hbar$ to 10 $\hbar$. Such fact confirms the saturation of population high-spin states for heavy frgment masses.

Figs. 2 and 3 show the variation of isomer ratios with the product mass lost from a hypothetical compound nucleus, $\Delta{A} = A_{cn} - A$ and with relative linear momentum transfer $p/p_{cn}$ determined in our recent work \cite{Balabekyan}, where $A_{cn}$ and $p_{cn}$ are the mass and momentum of a hypothetical compound nucleus and $A$ is the product number. There is a clear tendency of isomer ratio decrease with increase the linear momentum transfer (or $E^{*}$) in fast projectile-target interaction and the mass lost is observed. This is understandable since the probability of population high-spin state is associated with high angular momuntum transfer to the target.

From the Table I one can see that given residual nuclei have been formed in several different processes having  different excitations and characterized by different linear momentum transferred to the target in the first abrasion step of the reaction ($p$). Angular momentum transfer $l$ is determined by impact parameter of collision $b$ and the linear momentum of the projectile. The linear momentum of the projectile can be absorbed completely in head-on interaction when the impact parameter $b \approx 0$. Such collision does not produce large angular momentum since $l = p\cdot b$. The value corresponds to the transfer of a total linear momentum defines by the sum $b_{max} = (R_{p}+R_{T})$, where $R_{p}$ and $R_{T}$ are the radii of projectile target, respectively. The probability of such transfer is low because the momentum transfer at high energies is much low than total momentum of the projectile \cite{Saint-Laurent}. 

According to our result based on the estimation of total reaction cross section \cite{Balabekyan}, the impact parameter in deuteron-induced reaction of gold at 4.4 GeV is equals 8.37 fm, which corresponds to peripheral collision. As it was shown in \cite{Balabekyan}, the maximum linear momentum ($p/p_{cn} \sim 0.47$) can be released in the case of the production light nuclei with masses $A < 40$ probably in multifragmentation process of gold. It means that at the impact of 4.4 GeV deuterons with gold target a maximum angular momentum of about 80 $\hbar$ can be imparted. $^{44}$Sc isotope represents the nucleus of the intermediate mass fragments (IMFs) ($40 < A < 60$). Such fragments could be associate with fragmentation of the gold although fission, where the counter part heavy fragments would be in the mass region of $A \sim$ 120-130, also can give a contribution in formation of such isomeric pair. In the case of IMFs the average fractional linear momentum $\sim$ 0.16 and $<l> \sim$ 27 $\hbar$.
Niclides as $^{95}$Nb and $^{95}$Tc ($60 < A < 120$ amu) can be accosiate with fission-like and/or deep spallation processes. The average fractional momentum transfer in such range of product masses is $\sim$ 0.12 with $<l> \sim$ 20 $\hbar$. Although fission in the case of Au target is definitely less probable than deep spallation.

And the heavy mass fragments ($A > 131$ amu) as $^{184}$Re if the fragment produced by spallation of gold. Here the products have the lowest value of transfer excitation energy and linear momentum ($p/p_{cn} \sim 0.09$)  and angular momentum as well ($<l> \sim$ 15 $\hbar$).

It should be mentioned also that the linear momentum and correspondently excitation energy of fragments are proportional to the emitted nucleons. Thus, the spallation products are produced in peripheral collisions with large impact parameter. From the experimental determined fragment production cross sections \cite{Balabekyan} it was shown that the majority part of total reaction cross section produces by spallation, deep spallation and fission-like processes. Such fact is proved by the isomer ratio measurements of the present paper from the following point of view.
One can see from the Fig. 1 that isomer-to-ground state ratio is below unity for nucleus probably arise from fragmentation ($^{44}$Sc) and exceed unity in the range of plateau, for instance for $^{95}$Nb or $^{95}$Tc even at the same spin difference. It is means that at such energy of projectile the population of high-spin isomeric states prevail for reaction products formed mainly in fission-like/deep spallation and spallation processes and take place at the same excitation energy regime. It can be suggested that fragmentation requires more excitation energy for the formation with higher probability and contribution to the total reaction yield.

The population of the ground and metastable states of the residual $^{198g,m}$Au nuclei can be produced as independent cross sections directly from the (d, p) transfer reaction when the proton in deuteron nucleus escapes without further interaction. This may be explained as the incident neutron generating from a deuteron-nucleus collision interacts only at the periphery of the nucleus. Otherwise, in the deep-inelastic process the proton of the neutron-proton pair will be involved in an intranuclear cascade interaction through different reaction channel. Another possible way to produce $^{198g,m}$Au isomeric pair is through the reaction (n, $\gamma$) when nucleus has merely enough excitation energy so that nucleons cannot be emitted, but only $\gamma$-rays. There is evidence that deuteron as week bound system can decay and demonstrate one nucleon collision in the most cases of interactions \cite{Balabekyan}. 
The average angular momentum of the nucleus calculated in two above mentioned cases does not differ very much since the one proton, which is assumed to be emitted in the first case, does not take much angular momentum but take mainly
excitation energy: the average spin of the fragment is reduced by not more than 0.5 unit of angular momentum. 

The average initial angular momentum of the reaction remnant was deduced from the measured cross sections of $^{198}$Au isomer by calculations based on the concepts of the statistical model. The formalism of these calculations was introduced by Huizenga and Vandenbosch \cite{Huizenga} for the calculation of isomeric ratios in spallation reactions where the angular momentum distribution of the nuclei can be calculated. 

The probability distribution of initial angular momentum is assumed to be represented by following formula:

\begin{widetext}
\begin{eqnarray}
P(J_{i}) \sim (2J_{i}+1)exp[-J_{i}(J_{i}+1)/B^{2}],
\end{eqnarray}
\end{widetext}

\noindent 
where $P(J_{i})$ is the probability distribution for each spin value $J$, and $B$ is a parameter which defines the width of the distribution. The root-mean-square angular momentum $(\bar{J^{2}})^{1/2}$ of the primary product is equal to $B$ for large values of $B$:

\begin{widetext}
\begin{eqnarray}
(\bar{J^{2}})^{1/2} \cong B.
\end{eqnarray}
\end{widetext}

In this particular case the initial spin distribution is modified by the emission of the proton which carry away some angular momentum. 

In the reaction of 4.4 GeV deuterons with gold target the prefragment excitation energies $E^{*}$ were estimated on the base of linear momentum transfer of the projectile \cite{Balabekyan}. The effective excitation energy of residual nucleus of $^{198}$Au after emission of proton $E^{*}_{eff}$ was calculated using followinq expression:

\begin{widetext}
\begin{eqnarray}
E^{*}_{eff}=E^{*}-E_{Coul}-E_{p}-E_{KE}
\end{eqnarray}
\end{widetext}

\noindent where $E_{Coul}$ is the Coulomb barrier for the emitted proton, $E_{p}$ is the binding energy of a proton inside a nucleus, and $E_{KE}$ is the mean kinetic energy of proton. 

According to the evaporation model nucleons are emitted by the excited nucleus with a mean energy $E_{KE}=2T$, where $T$ is the nuclear temperature determined by the expression \cite{Blatt}:

\begin{widetext}
\begin{eqnarray}
aT^{2}-4T = E^{*}_{eff},
\end{eqnarray}
\end{widetext}

\noindent where parameter $a$ is the level-density parameter ($a=A/10$ MeV$^{-1}$).

At the stage of the process involving the cascade of $\gamma$-transitions eventually leading to the metastable or the ground state, the density of the nuclear level spin distribution determines the probability of population of intermediate nuclear states. In the calculations, we used the spin part of the Bethe-Bloch formula given by: 

\begin{widetext}
\begin{eqnarray}
P(J) \sim (2J+1)exp[-(J+0.5)^{2}/2\sigma^{2}],
\end{eqnarray}
\end{widetext}

\noindent where P(J) is the probability distribution of levels with spin $J$ and $\sigma$ is the spin cutoff parameter
which characterizes the angular momentum distribution of the level density and is related to the moment of inertia and the temperature of the excited nucleus and given by $\sigma^{2}=0.00889\sqrt{aE^{*}_{eff}}A^{2/3}$.

The average energy and number of prompt $\gamma$-rays emitted from the nucleus of initial excitation energy $E^{*}_{eff}$ can be estimated by means of the formula \cite{Huizenga1}:

\begin{widetext}
\begin{eqnarray}
\bar{E_{\gamma}}=4\left(\frac{E^{*}_{eff}}{a}-\frac{5}{a^{2}}\right)^{1/2}.
\end{eqnarray}
\end{widetext}

The energy of each succeeding $\gamma$-ray is found by computing the new excitation energy by subtracting from
the residual excitation energy the average energy of the $\gamma$-ray calculated by use of Eq. (9).

During calculations we used both dipole $E1$ and quadrupole $E2$ multipolity of $\gamma$-transitions. However, it was shown in \cite{Aumann} the relative share of the quadrupole transitions in the deexcitation process does not exceed 10\%.
 
The excitation energy $E^{*}_{eff}$ and, accordingly, the energy $E_{\gamma}$ of emitted photons are determined at
each stage of the cascade. The last level from which the population of the ground or the isomeric state occurs is
characterized by an excitation energy not higher than 2 MeV.

The results of calculations of the angular momentum parameter induced in the compound nucleus results of $^{198g,m}$Au nuclei was equal $B = 20.5\pm5.0 \hbar$. The uncertainty of the calculation have been compounded from the statistical errors in the activity measurements for isomer ratio and the uncertainty in our knowledge of the deexcitation cascade.
Thus, the average spin of compound nucleus after interaction of 4.4 GeV deuterons with $^{197}$Au target was considered to increase from 2.5 $\hbar$ to 20.5 $\hbar$. These result indicates that considerable intrinsic angular momentum can be produced in the spallation process.

\newpage
\begin{center}
Table I. The production cross sections for ground and metastable states and corresponding isomer rations of nuclides formed by the reaction of 4.4 GeV deuterons with ${}^{197}$Au. 
\end{center}
\begin{center}
\begin{tabular}{||c||c||c||c||c||} \hline\hline
Element & State/Spin & $\sigma_{m}$, mb & $\sigma_{g}$, mb &$\sigma(I_{h})/\sigma(I_{l})$\\
\hline $^{44}$Sc& m(6$^{+}$) & 0.45$\pm$0.16 & 1.76$\pm$0.70 & 0.26$\pm$0.09\\ 
 & g(2$^{+}$) &  & & \\
\hline $^{95}$Nb& m(1/2$^{-}$) & 0.35$\pm$0.03 & 1.09$\pm$0.30 & 3.11$\pm$0.87\\ 
 & g(9/2$^{+}$) &  & & \\
 \hline $^{95}$Tc& m(1/2$^{-}$) &  &  & 3.11$\pm$0.87\\ 
 & g(9/2$^{+}$) &  & & \\
\hline $^{184}$Re& m(8$^{+}$) & 4.09$\pm$0.25 & 0.92$\pm$0.18 & 4.45$\pm$1.11\\ 
 & g(3$^{-}$) &  & & \\
\hline $^{198}$Au& m(12$^{-}$) & 7.41$\pm$0.50 & 1.56$\pm$0.25 & 4.75$\pm$1.05\\ 
 & g(2$^{-}$) &  & &\\ \hline \hline
\end{tabular}
\end{center}
\vspace{2cm}

\newpage
\begin{figure*}[h!]
\includegraphics[width=16cm]{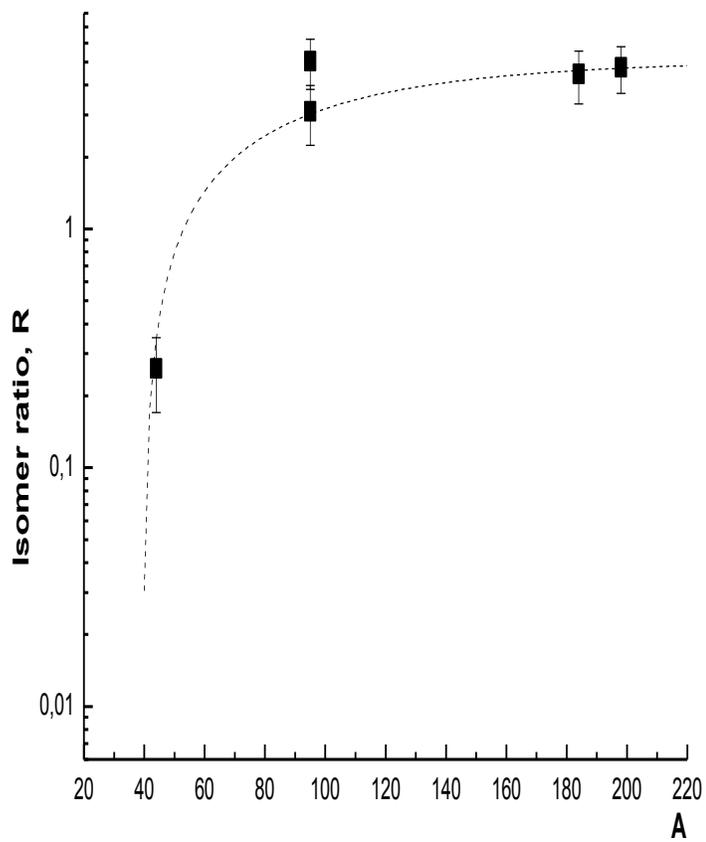}
\caption{\small Isomer ratios, $IR$, versus product mass number $A$. The dashed line shows the general trend of $IR$.} 
\end{figure*}

\newpage
\begin{figure*}[h!]
\includegraphics[width=16cm]{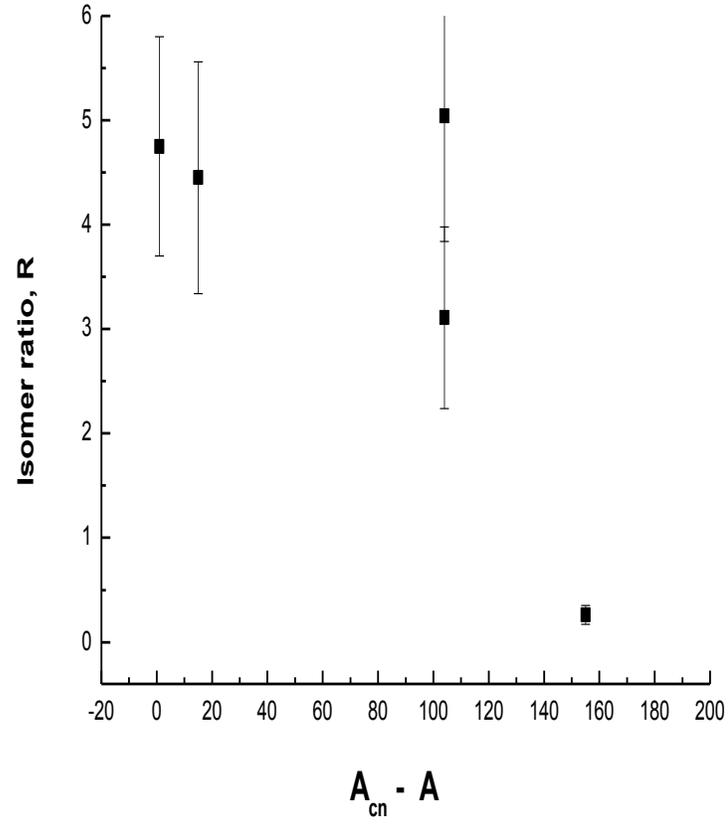}
\caption{\small Dependence of the isomer ratios, $IR$, on the product mass losses from a hypothetical compound nucleus formed in a complete fusion $\Delta{A} = A_{cn} - A_{pr}$.}
\end{figure*}

\newpage
\begin{figure*}[h!]
\includegraphics[width=16cm]{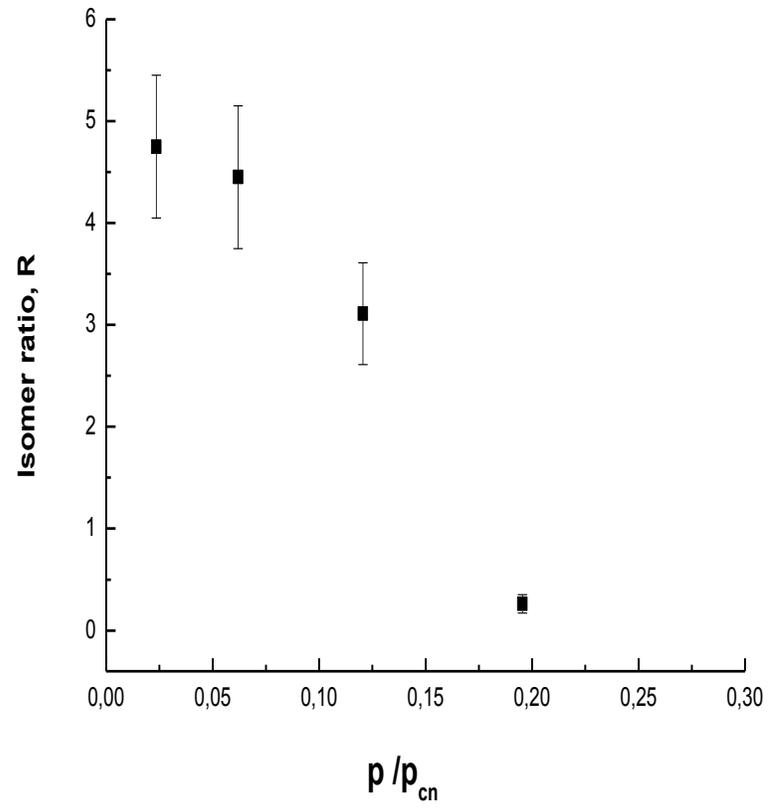}
\caption{\small Dependence of the isomer ratios, $IR$, on the the fractional momentum transfer of residual cascade nuclei  $p/p_{cn}$ from \cite{Balabekyan}.}
\end{figure*}

\section*{4 Conclusion}
In the present paper for the first time up to our knowledge a number of isomer ratio values are determined in order to investigate general regularities of this phenomenon for deuteron-induced reactions. $^{197}$Au targets were bombarded by 4 GeV to produce a broad variety of isomers with the high spin values. Based on experimental results, a qualitative explanation of the observed regularities is suggested. It was found that there is a competition of different processes during the population of metastable or ground states and the main contribution goes from the spallation/fission reaction channels.
The proposed reaction mechanism may include also the emission fragments through decay of high-excited gold target. 
The average angular momenta of compound nucleus calculated in the framework of the statistical-model analysis was extracted. It was concluded that at the interaction of high-energy deuterons with gold target average spin of compound nucleus increases from 2.5 $\hbar$ to 20.5 $\hbar$.

\section*{Acknowledgment}
G. Karapetyan is grateful to Funda\c c\~ao de Amparo \`a Pesquisa do Estado de S\~ao Paulo (FAPESP) 2011/00314-0 and  to International Centre for Theoretical Physics (ICTP) under the Associate Grant Scheme.

\medbreak

\bigskip

\begin{thebibliography}{99}
\bibitem{Karamian} S. A. Karamian and J. J. Carroll, Phys. Rev. C $\bf83$, 024604 (2011).{} 
\bibitem{Zhuikov} B. L. Zhuikov, M. V. Mebel, V. M. Kokhanyuk $\textit{et al.}$, Phys. Rev. C $\bf68$, 054611 (2003).{}
\bibitem{Patronis} N. Patronis, C. T. Papadopoulos, S. Galanopoulos $\textit{et al.}$, Phys. Rev. C $\bf16$, 034607 (2007).{}
\bibitem{Demekhina} N. A. Demekhina, A. S. Danagulyan, and G. S. Karapetyan, Phys. At. Nucl. $\bf65$, 365 (2002).{}
\bibitem{Orth} C. J. Orth $\textit{et al.}$, Phys. Rev. C $\bf21$, 2524 (1980).{}
\bibitem{Aumann} D. C. Aumann $\textit{et al.}$, Phys. Rev. C $\bf16$, 254 (1977).{}
\bibitem{Firestone} R. B. Firestone, in Tables of Isotopes, 8th ed.: 1998 Update (with CD ROM), edited by S. Y. Frank Chu (CD-ROM editor) and C. M. Baglin (Wiley Interscience, New York, 1996).{}
\bibitem{Balabekyan} A. R. Balabekyan, N. A. Demekhina, G.S. Karapetyan $\textit{et al.}$, submitted to PRC.{}
\bibitem{Vanska} R. Vanska and R. Rieppo, Nucl. Instrum. Methods $\bf179$, 525 (1981).{}
\bibitem{Kolev} D. Kolev, E. Dobreva, N. Nenov, and V. Todorov, Nucl. Instrum. Methods $\bf356$, 390 (1995).{}
\bibitem{Deppman} A. Deppman $\textit{et al.}$, Phys. Rev. C $\bf21$, 2524 (2013).{}
\bibitem{Hufner} J. Hufner, Phys. Rep. $\bf125$, 129 (1985).{}
\bibitem{Saint-Laurent} F. Saint-Laurent, M Conjeaud, R. Dayras $\textit{et al.}$, Nucl. Phys. A $\bf422$, 307 (1984).{}
\bibitem{Huizenga} J. R. Huizenga and R. Vandenbosch, Phys. Rev. $\bf120$, 1305 (1960).{}
\bibitem{Blatt} J. M. Blatt and V. F.Weisskopf, Theoretical Nuclear Physics (Wiley, New York, 1952; Inostrannaya Literatura, Moscow, 1954).{}
\bibitem{Huizenga1} H. K. Vonach, R. Vandenbosch, and J. R. Huizenga, Nucl. Phys. $\bf60$, 70 (1964).{}
\end{thebibliography}
\end{document}